\documentstyle[aps,prl,twocolumn]{revtex}

\begin{document}
\draft

%\bibliographystyle{alpha.bst}
%\bibliography{ref.bib}

\title{From the magnetic-field-driven transitions to the 
zero-field transition in two-dimensions}

\author{Y. Hanein, D. Shahar and Hadas Shtrikman}

\address{ Dept. of Condensed Matter Physics, Weizmann Institute, 
Rehovot 76100, Israel}

\author{J. Yoon, C.C. Li and D.C. Tsui}

\address{Dept. of Electrical Engineering, Princeton University, 
Princeton, New Jersey 08544}

\maketitle

 % \begin{abstract}
 % \end{abstract}
 % 
 % \pacs{71.30.+h,73.40.-c}

\textbf{
For more than a decade it was widely accepted that
two-dimensional electrons are insulating at zero 
temperature and at zero magnetic-field \cite{Abrahams1979}. 
Experimentally it was demonstrated \cite{Jiang93} that, when placed in a strong 
perpendicular magnetic field,
the insulating phase turns into a quantum-Hall state.
While this transition was in accordance with existing theoretical 
models \cite{KLZ}, the density driven	metal-insulator	transition at zero magnetic-field, 
recently observed
in high-quality two-dimensional systems \cite{SVKrav94}, 
was unforeseen and, despite 
considerable amount of effort, its origins are still unknown
\cite{SFinkel84,VDobrosavljevic97,Maslov,PPhillips98,SongHe}.
In order to improve our understanding of the zero magnetic-field transition, 
we conducted a study of
the insulator to quantum-Hall transition in low-density, 
two-dimensional, hole system in GaAs that exhibits the zero magnetic-field 
metal-insulator transition \cite{Hanein1998}. 
We found that, in the low field insulating phase, upon increasing the carrier 
density towards the metal-insulator transition, the 
critical magnetic-field of the insulator to quantum-Hall transition decreases and 
converges to the zero magnetic-field metal-insulator transition. 
This implies a common origin for both the 
finite magnetic-field and the zero magnetic-field transitions.} \\

In Fig. 1a we plot the resistivity ($\rho$) of one of our samples as a function 
of magnetic-field ($B$) at several temperatures, with the hole-density 
($p$) held fixed .
At $B=0$ the system is insulating as indicated by a  rapidly increasing
$\rho$ as the temperature ($T$) approaches zero. 
The insulating behavior is maintained for 
$B<B_{c}^{L}$.
For $B>B_{c}^{L}$, a quantum-Hall (QH) state ($\nu=1$, where $\nu$ is 
the Landau-level filling factor) is observed with $\rho$ 
tending to 
zero upon lowering of $T$. 
We identify $B_{c}^{L}$, the point where the temperature coefficient of 
resistivity (TCR) changes its sign, with the critical point of the 
insulator-to-QH transition \cite{Jiang93}. 
At still higher $B$, beyond the $\nu=1$ QH state, 
the system turns insulating and a second 
$T$-independent transition 
point is seen at $B_{c}^{H}$.
$B^{H}_{c}$ therefore marks the critical $B$ of the  QH-to-insulator 
transition \cite{universality}.  
Following the path set by earlier studies \cite{Shashkin,Jiang}  we
focus, for now, on the low-$B$ 
transition 
and follow, in Figs. 1b-1e, the evolution of the critical point $B_{c}^{L}$ 
as we increase $p$.  \\

Data obtained from the 
same sample at successive increases of the density are shown in Figs. 1b-1e. 
As in Fig. 1a, the insulator-to-QH transition point is 
evident in Fig 1b, but the transition point ``moves'' to a lower $B$. 
This trend continues in Fig. 1c until finally, in Fig. 1d, the crossing point
disappears. 
Along with this shift in $B^{L}_{c}$, 
we notice in Figs. 1a-1c, that the insulating behavior at $B=0$ becomes weaker 
until, in Fig. 1d, $\rho$ at $B=0$ is $T$-independent. 
In the next graph, Fig. 1e, the system has crossed over into
its metallic phase and no transition, or $T$-independent point, is seen
implying that the density of Fig. 1d ($p=1.34\cdot 10^{10}$ 
cm$^{-2}$) is the  critical density of the 
metal-insulator 
transition (MIT) at $B=0$. 
This $B=0$ transition is the  MIT in two dimensions (2D) first reported by 
Kravchenko \emph{et al.} for Si samples \cite{SVKrav94}. 
Our main result can now be stated: Upon increasing $p$, 
$B^{L}_{c}$ gradually tends to lower $B$'s, eventually converging to 
the $B$=0 MIT which, for this sample, takes place at $p=1.34\cdot 10^{10}$ 
cm$^{-2}$.\\

To complement our $p$-dependence study of the $B$-driven transitions
we will next focus on the effect of a perpendicular $B$ on the $p$-driven transition.
Our new starting point is the more conventional experimental demonstration of the $B=0$ 
MIT in 2D. In Fig. 2a we plot $\rho$ as a function 
of $p$ at 
several $T$'s and at $B$=0. A $T$-independent crossing point is seen 
here as well (at $p_{c}=1.34\cdot 10^{10}$ cm$^{-2}$), marking the transition 
from insulating behavior for 
$p<p_{c}$ to metallic behavior for $p>p_{c}$.
We then repeat, in Figs. 2b-2e, the measurement of Fig. 2a at different 
values of $B$.
In Figs. 2b and 2c, $p_{c}$ shifts to a lower value, a trend which 
reverses for $B\geq 0.35$T (Figs. 2d-2e). This trend reversal 
of $p_{c}(B)$ is accompanied by the development of non-monotonic 
dependence of $\rho$ on $p$, which is a precursor to the quantum Hall 
effect (QHE). We now 
combine these $p_{c}(B)$ results with the $B_{c}(p)$ of Fig. 1, to plot a 
comprehensive phase diagram of our system in the $B-p$ plane. \\

The phase diagram obtained from our
data is shown in Fig. 3,
where we plot the $B$ and $p$ coordinates of each one of the 
transitions. Separate symbols are given to $B_{c}^{L}$ and $B_{c}^{H}$, 
defined in Fig. 1, and to $p_{c}$ from Fig. 2. Several points emerge from 
inspecting the resulting phase diagram. First, 
we note that the results obtained from the two data sets (fixed $p$ 
and fixed $B$ measurements) are mutually consistent. 
Second, for fixed $p$'s between $0.88$ and $1.33\cdot 10^{10}$ cm$^{-2}$
the low-$B$ insulating phase first turns metallic and then reappears at 
high $B$.   
This reentrant nature of the insulating phase is clearly reflected in 
the $B$ traces of Figs. 1a-1c.
And third, the low $B$ region of the phase boundary 
reiterates the main result of our work and clearly depicts the 
continuous evolution of the transition from high-$B$ to the $B$=0 
MIT.  The relation between the transition at finite $B$ and the 
MIT transition at $B$=0,
suggests that similar processes govern the transport for both 
transitions \cite{Shimshoni}. \\

Fig. 3 also includes the high-$B$ side of the phase diagram 
($B^{H}_{c}$). In fact, the low and high $B$ regimes are smoothly	
connected to form a single phase-boundary line. 
It is common practice to describe the finite-$B$ transitions in the 
language of quantum phase transitions \cite{Sondhi}.
If we assume that 
the phase-boundary line of Fig. 3 comprises a set of quantum critical points, 
it is possible that 
universal features should be observed in its vicinity.
To test this proposition we examine the value of $\rho$ 
at the transition points, $\rho_{c}$.
In Fig. 4a we plot $\rho_{c}$ of our transitions as a function of $B$.
Overall, $\rho_{c}$ is not constant, its value changing by 
almost a factor of 4 over our $B$ range.
However, at very low as well as very high $B$, $\rho_{c}$ 
approaches a value close to $h/e^{2}$, the quantum unit of resistance. 
Although for our sample, at $B$=0, $\rho_{c}$  is close to $h/e^{2}$ 
we wish to point out that the value of  $\rho_{c}$ at $B$=0 obtained 
from other samples 
varies by an order of magnitude, between 0.4 and 4$h/e^{2}$ \cite{Hanein2deg}.  \\

So far we have shown that $B^{L}_{c}$, $B^{H}_{c}$ and $p_{c}$ define 
a common phase boundary line in the $B-p$ plane, and that at the 
intermediate $B$ range 
along this phase boundary $\rho_{c}$ significantly deviates from 
$h/e^{2}$, its value near $B$=0 and at high-$B$.
It is instructive to consider the dependence of $\rho_{c}$ on $p$, 
rather than on $B$. 
In Figs. 1a-1c we can readily see the general trend: $\rho_{c}$ of the 
low and high-$B$ transitions at fixed $p$ are very close to 
each other. 
To test this result for our entire range of $p$ we plot, in Fig. 4b,
$\rho_{c}$ versus $p$ obtained from 
our data.
We see that the two 
transitions have collapsed onto a single curve for 
our entire range of $p$.
This result demonstrates that for a given carrier-density $\rho_{c}$ of the 
low and high-$B$ transitions is the same. This supports the 
notion of symmetry between these transitions \cite{Hilke}.  \\

A possible relation between different transitions in 2D systems 
was noted by Jiang \emph{et al.} \cite{Jiang93} who 
pointed out the similarities between the insulator-to-QH 
and the insulator-to-superconductor transitions.
Theoretical basis for such similarity was introduced in ref. \cite{Shimshoni}.
In our work we found a relation between 
the finite $B$ insulator-to-QH transition and the 
metal-insulator transition at $B$=0.
Since both transitions are 
measured in the same 2D system, we were able to continuously transform 
one to the other.
This raises the possibility that both transitions share a common 
physical origin. \\

\begin{flushleft}
	\textbf{Methods}
\end{flushleft}
The sample used in this study is a $p$-type, 
inverted semiconductor insulator semiconductor (ISIS) sample 
\cite{Meirav1988} grown on (311)A GaAs substrate
\cite{Hanein1997} with Si as a $p$-type dopant.
In an ISIS device the carriers are
accumulated in an undoped GaAs layer lying on top of an undoped AlAs barrier,
grown over a $p^{+}$ conducting layer. This $p^{+}$ conducting layer 
is separately contacted and serves as a back-gate.
The hole carrier-density ($p$) is varied by applying voltage ($V_{g}$) 
to the back-gate. 
The sample was wet-etched to the shape of a
standard Hall-bar and the 
measurements were done in a dilution refrigerator with a base $T$ of 
$57$ mK, using AC lock-in technique with an excitation current of $1$ nA
flowing in the
[01$\bar{1}$] direction. \\

\begin{flushleft}
	\textbf{Acknowledgments.}
\end{flushleft}
The authors wish to thank Efrat Shinshoni, M. Hilke and Amir Yacoby for very 
interesting discussions.
This work was supported by the NSF, the BSF and by a grant from the Israeli 
Ministry of Science and The Arts.  \\

\begin{figure}  
\caption{Isotherms of the magneto-resistance data of sample H324Bc at 
various densities.
The traces are color coded with blue being low and red being high 
$T$.
$T$=63, 122, 177, and 217 mK
(a) $p=1.09\cdot10^{10}$ cm$^{-2}$.
(b) $p=1.21\cdot10^{10}$ cm$^{-2}$.
(c) $p=1.26\cdot10^{10}$ cm$^{-2}$.
(d) $p=1.34\cdot10^{10}$ cm$^{-2}$.
(e) $p=1.42\cdot10^{10}$ cm$^{-2}$.}
\end{figure}

\begin{figure}  
\caption{Isotherms of the resistance versus $p$ of sample H324Bc at 
various $B$'s.
The traces are color coded with blue being low and red being high 
$T$. 
(a) $B$=0 T at $T$=57, 120, 160, and 214 mK.
(b) $B$=0.2 T at $T$=58, 121, 160, and 217 mK.
(c) $B$=0.4 T at $T$=60, 98, 148, and 212 mK. 
(d) $B$=0.6 T at $T$=59, 122, 161, and 217 mK.
(e) $B$=0.9 T at $T$=59, 122, 161, and 217 mK.}
\end{figure}

\begin{figure}  
\caption{
$B$ and $p$ coordinates of the crossing points derived from data similar 
to those of Figs. 1a-1e and Figs. 2a-2e.
Open diamonds denote $p_{c}$, solid diamonds denote $B^{L}_{c}$
and solid triangles denote $B^{H}_{c}$.
The arrow points to the $B$=0 MIT.}
\end{figure}

\begin{figure}  
\caption{
(a) $\rho_{c}$ plotted as a function of $B$.
Open diamonds denote  $\rho_{c}$ from $p_{c}$ data, 
solid diamonds denote $B^{L}_{c}$ 
data and solid triangles denote $B^{H}_{c}$ data.
(b) $\rho_{c}$ plotted as a function of $p$. }
\end{figure}


\begin{references}

\bibitem{Abrahams1979}  
Abrahams, E., Anderson, P. W., Licciardello, D. C. \& Ramakrishnan, 
T. V.
Scaling theory of localization: Absence of quantum diffusion in two 
dimensions. 
Phys. Rev. Lett. {\bf 42}, 673-676 (1979). %%

\bibitem{Jiang93} 
Jiang, H. W., Johnson, C. E., Wang, K. L. \& Hannahs, S. T.
Observation of magnetic-field-induced delocalization- 
transition from Anderson insulator to quantum Hall conductor.
Phys. Rev. Lett. {\bf 71}, 1439-1442 (1993). %%

\bibitem{KLZ}
Kivelson, S., Lee, D. H. \& Zhang, S. C.
Global phase diagram in the quantum Hall effect.
Phys. Rev. B {\bf 46}, 2223-2238 (1992). 

\bibitem{SVKrav94}  
Kravchenko, S. V., Kravchenko, G. V., Furneaux, J. E., Pudalov, V. 
M. \& D'Iorio, M.
Possible metal-insulator transition at $B$=0 in two dimensions.
Phys. Rev. B {\bf 50}, 8039-8042 (1994). %%

\bibitem{SFinkel84} 
FinkelÕstein, A. M. 
Influence of Coulomb interaction on the properties of disordered metals.
Zh. Eksp. Teor. Fiz. {\bf 84}, 168-189 (1983) [JETP {\bf 57},
97-108 (1983)].

\bibitem{VDobrosavljevic97} 
Dobrosavljevic, V., Abrahams, E., Miranda, E. \& Chakravarty, S.
Scaling theory of two-dimensional metal-insulator transitions.
Phys. Rev. Lett. {\bf 79}, 455-458 (1997).

\bibitem{Maslov}
Altshuler, B. L. \&  Maslov D. L. 
Theory of Metal-Insulator Transitions in Gated Semiconductors.
Phys. Rev. Lett. {\bf 82}, 145-148  (1999)
 
\bibitem{PPhillips98}
Phillips, P., Wan, Y. I., Martin, I., Knysh, S. \& Dalidovich, D. 
Superconductivity in a two-dimensional electron gas.
Nature {\bf 395}, 253-257 (1998)

\bibitem{SongHe}
He, S. \& Xie, X.C.
New liquid phase and metal-insulator transition in Si-MOSFETs.
Phys. Rev. Lett. {\bf 80}, 3324-3327 (1998). %%

\bibitem{Hanein1998}
Hanein, Y., Meirav, U., Shahar, D., Li, C.C., Tsui, D.C. \& Shtrikman H. 
The metalliclike conductivity of a two-dimensional hole system.
Phys. Rev. Lett. {\bf 80}, 1288-1291 (1998). %%

\bibitem{universality}
Shahar, D., Tsui, D. C., Shayegan, M., Bhatt, R. N. \&  Cunningham, J. E. 
Universal conductivity at the quantum Hall liquid to insulator
transition.
Phys. Rev. Lett. {\bf 74}, 4511-4514 (1995).


\bibitem{Shashkin}
Shashkin, A. A., Dolgopolov, V. T. \& Kravchenko, G. V.
Insulating phases in a two-dimensional electron system of high 
mobility Si MOSFET's.
Phys. Rev. B. {\bf 49}, 14486-14495 (1994). %%

\bibitem{Jiang} 
Dultz, S. C., Jiang, H. W. \& Schaff, W. J.
Absence of floating delocalization states in a
two-dimensional hole gas.
Phys. Rev. B {\bf 58}, R7532-R7535 (1998).

\bibitem{Shimshoni}
Shimshoni, E., Auerbach A. \& and Kapitulnik	A.
Transport through Quantum Melts.	   
Phys. Rev. Lett.	{\bf 80}, 3352-3355	(1998).

\bibitem{Sondhi}
Sondhi, S. L., Girvin, S. M., Carini, J. P. \& Shahar, D.
Continuous quantum phase transitions.
Rev. Mod. Phys. {\bf 69}, 315-333 (1997).

\bibitem{Hanein2deg}
Hanein, Y.,  Shahar, D., Yoon, J.,  Li, C.C., Tsui, D.C. \& Shtrikman H. 
Observation of the metal-insulator transition in
two-dimensional n-type GaAs.
Phys. Rev. B {\bf 58}, R13338-R13340 (1998).

\bibitem{Hilke}
Hilke, M., Shahar, D., Song, S. H., Tsui, D. C., Xie, Y. H. \& Monroe, D.
Symmetry in the insulator-Hall-insulator transitions observed in a 
Ge/SiGe quantum well.
Phys. Rev. B. {\bf 56}, 15545-15548 (1997).%%

%% 
 % \bibitem{DShahar95}
 % Shahar, D., Tsui, D.	C. \& Cunningham, J. E.
 % Observation of the $\nu=1$ quantum Hall effect in a strongly	localized 
 % tow-dimensional system.	 
 % Phys. Rev. B. {\bf 74}, 14372-14375 (1995). %%
 %%


\bibitem{Meirav1988}  
Meirav, U., Heiblum, M. \& Stern, F.
High-mobility variable-density two-dimensional electron gas in 
inverted GaAs-AlGaAs hetrojunctions.
Appl. Phys. Lett. {\bf 52}, 1268-1270 (1988)

\bibitem{Hanein1997}  
Hanein, Y., Shtrikman, H. \& Meirav, U.
Very low density two-dimensional hole gas in an inverted GaAs/AlAs interface. 
Appl. Phys.
Lett. {\bf 70}, 1426-1428 (1997). %%
 


\end{references}
\end{document}